# Density functional modeling of the binding energies between aluminosilicate oligomers and different metal cations

[1,2]Kai Gong, [1]Kengran Yang, [1]Claire E. White

[1]Department of Civil and Environmental Engineering, Andlinger Center for Energy and the Environment, Princeton University, New Jersey 08544, USA

[2]Department of Civil and Environmental Engineering, Rice University, Houston, Texas 77005, USA (current address)

**Abstract**

Interactions between negatively charged aluminosilicate species and positively charged metal cations are critical to many important engineering processes and applications, including sustainable cements and aluminosilicate glasses. In an effort to probe these interactions, here we have calculated the pair-wise interaction energies (i.e., binding energies) between aluminosilicate dimer/trimer and 17 different metal cations $M^{n+}$ ($M^{n+}$ = $Li^+$, $Na^+$, $K^+$, $Cu^+$, $Cu^{2+}$, $Co^{2+}$, $Zn^{2+}$, $Ni^{2+}$, $Mg^{2+}$, $Ca^{2+}$, $Ti^{2+}$, $Fe^{2+}$, $Fe^{3+}$, $Co^{3+}$, $Cr^{3+}$, $Ti^{4+}$ and $Cr^{6+}$) using a density functional theory (DFT) approach. Analysis of the DFT-optimized structural representations for the clusters (dimer/trimer + $M^{n+}$) shows that their structural attributes (e.g., interatomic distances) are generally consistent with literature observations on aluminosilicate glasses. The DFT-derived binding energies are seen to vary considerably depending on the type of cations (i.e., charge and ionic radii) and aluminosilicate species (i.e., dimer or trimer). A survey of the literature reveals that the difference in the calculated binding energies between different $M^{n+}$ can be used to explain many literature observations associated with the impact of metal cations on materials properties (e.g., glass corrosion, mineral dissolution, and ionic transport). Analysis of all the DFT-derived binding energies reveals that the correlation between these energy values and the ionic potential and field strength of the metal cations are well captured by 2nd order polynomial functions ($R^2$ values of 0.99-1.00 are achieved for regressions). Given that the ionic potential and field strength of a given metal cation can be readily estimated using well-tabulated ionic radii available in the literature, these simple polynomial functions would enable rapid estimation of the binding energies of a much wider range of cations with the aluminosilicate dimer/trimer, providing guidance on the design and optimization of sustainable cements and aluminosilicate glasses and their associated applications.

# 1 Introduction

The interactions between aluminosilicates and metal cations are important for many engineering processes and applications. One example is the formation of cementitious materials, including alkali-activated materials (AAMs) and blended cements, which bind aggregates together to form concrete. AAM is an important sustainable material technology that is able to convert a solid precursor source (e.g., industrial wastes and calcined clay rich in amorphous aluminosilicates) to a cementitious binder [1]. The final AAM binder has many potential applications, including being used as a low-$CO_2$ cement alternative to Portland cement (PC) [2, 3], whose production worldwide is currently responsible for ~8% of global anthropogenic $CO_2$ emissions [4]. For geopolymers, i.e., AAMs based on low-Ca precursors (e.g., metakaolin and class F fly ash), the main binder gel responsible for most of its engineering properties is an amorphous three-dimensional alkali-alumino-silicate-hydrate (N-A-S(-H), when Na is the alkali) gel mainly consisting of $Q^4(m\text{Al})$ ($m$ = 0, 1, 2, 3, 4) for the silica units [3]. Alkali cations, for example, $Na^+$ and $K^+$, charge-balance the negatively charged alumina tetrahedra $(Al(O_{1/2})_4)^{-1}$, thereby stabilizing the aluminosilicate network. This charge-balancing interaction (interaction between negatively charged alumina tetrahedra and metal cations) helps to hinder ionic transport in calcium-alumino-silicate-hydrate (C-(A)-S-H, compared with calcium-silicate-hydrate (C-S-H)) gel [5], stabilize zeolite framework structures [6] and reduce alkali leaching from geopolymer binders [3]. Furthermore, this interaction in geopolymers has been used to immobilize heavy metals [7, 8] and treat wastewater [9]. On the other hand, excess alkali metal cations beyond those needed for charge-balancing act as modifiers to depolymerize the aluminosilicate network structure, as has been shown recently for sodium-substituted calcium-alumino-silicate-hydrate (C-(N)-A-S-H) [10].

The negatively charged aluminosilicate network can also be charge-balanced by alkaline earth metal cations (e.g., $Ca^{2+}$ and $Mg^{2+}$). In addition to charge-balancing, these alkaline earth metal cations, beyond those required for charge balancing, are also effective network modifiers, causing the aluminosilicate gel network to depolymerize. The resulting binder gels (e.g., C-A-S-H or magnesium-alumino-silicate-hydrate (M-A-S-H)) possess different atomic structures (mainly short aluminosilicate chain structure ($Q^2$) for C-(A)-S-H [11] and plane structure ($Q^3$) for M-A-S-H [12]), pore structures [13-16], mechanical properties [17], transport properties [14, 15, 18, 19] and chemical stability [15, 18]), compared with the three-dimensional N-A-S(-H) gel.

Another important example where the interactions between metal cations and aluminosilicates are critical is aluminosilicate glass containing alkali and/or alkaline earth metal cations. Aluminosilicate glasses are ubiquitous in many important industrial applications, including nuclear waste encapsulation, high-performance glasses, ceramics, metallurgical processes, and sustainable cement [20-23]. The metal cations in these aluminosilicate glasses play two distinct roles: (i) to charge-balance the negatively charged alumina tetrahedra (i.e., $(Al(O_{1/2})_4)^{-1}$), and (ii) to depolymerize the aluminosilicate network creating non-bridging oxygen (NBO) atoms (i.e., oxygen atoms bonded to only one silica or alumina tetrahedra). Many studies have shown that the type of alkali and alkaline earth metal cations has a dramatic impact on the resulting glasses, affecting both the atomic structure [22, 24-27] and engineering properties (e.g., physical [28, 29], mechanical [28, 30], thermal [28, 31] and chemical properties [32-34]). Furthermore, the type of metal cations also has a significant impact on the dissolution of silicate minerals and glasses [22, 35], which is critical to soil fertility, transport and sequestration of contaminants, and global geochemical cycle (including the $CO_2$ cycle) [35-37].

However, fundamental studies on the atomic scale interactions between different metal cations and aluminosilicate networks are limited since these detailed interactions are often difficult to elucidate using experiments. For this purpose, atomistic simulations are ideal for simulating their interactions. Recently, we have performed density functional theory (DFT) calculations to determine the pair-wise interaction energies (Gibbs free energies) between different monomeric species (silicate, aluminate, sodium, and calcium ions), gaining insight into the early stage formation mechanisms of different binder gels (i.e., C-S-H, C-A-S-H, and C-(N)-A-S-H gels), where the gels are responsible for most of the engineering properties of modern concrete [38]. Previously, a similar computational framework has been adopted to calculate the interaction energies between silicate and aluminate species [39], which, when combined with a coarse-grained Monte Carlo (CGMC) model, enables quantitative modeling of the early stages of formation of N-A-S(-H) gel in geopolymers [40, 41]. Similar DFT approaches have been adopted to calculate interaction energies among silicate species [42], silicate and alkali species [43, 44], and phosphate species [45]. They have also been used to calculate interaction energies between different metal cations (e.g., $Li^+$, $Na^+$, $K^+$, $Ca^{2+}$, and $Mg^{2+}$) and zeolitic frameworks [6] and organic matter (e.g., guanine and 6-thioguanine tetrads [46], glutathione [47], tetraoxa[8]circulene sheet [48], and cubane, cyclohexane and adamantane [49]).

A survey of the literature reveals that few computational studies have investigated the interactions between aluminosilicates and different metal cations, in spite of their prevalence in many important applications, as briefly mentioned above. In this study, we have probed the pair-wise interactions between aluminosilicate dimer/trimer and over seventeen metal cations/clusters (e.g., $Li^+$, $Na^+$, $K^+$, $Cu^+$, $Cu^{2+}$, $Co^{2+}$, $Zn^{2+}$, $Ni^{2+}$, $Mg^{2+}$, $Ca^{2+}$, $Ti^{2+}$, $Fe^{2+}$, $Fe^{3+}$, $Co^{3+}$, $Cr^{3+}$, $Ti^{4+}$ and $Cr^{6+}$) that are relevant to the aforementioned applications using DFT calculations. Detailed analysis of the DFT-optimized structures has been carried out to determine their interatomic distances, which were compared with literature data on silicate glasses, minerals, or clusters, to ensure that the cluster structures obtained are reasonable. Their pair-wise interaction energies (or binding energies) were determined and compared in the context of existing literature, where the observed trends in the interaction energies of different cations have been correlated with different literature observations (including the impact of different metal cations on glass corrosion, mineral dissolution, and ionic transport). Finally, the correlations between interaction energies and the attributes of different metal cations (e.g., charge, ionic radii, ionic potential, and field strength) have been explored to identify simple empirical equations to enable rapid estimation of interaction energies for unexplored cations. Although we are limited to pair-wise interaction (as opposed to simulating large cluster reactions) due to computation cost, this investigation illustrates the value of using simple model systems to better understand the impact of metal cations on the properties of aluminosilicate materials, which is critical to many important applications.

## 2 Methodology

Density functional theory calculations were performed to estimate the pair-wise interaction energies between aluminosilicate dimers and trimers and different charge balancing cations (see Table 1 for all the studied pairs), following procedures similar to our previous studies [38, 39]. Specifically, for each individual species (e.g., "dimer", "trimer", and "dimer/trimer + cation"), we first performed simulated annealing on gas-phase clusters using *ab initio* molecular dynamics (MD) simulations at different temperatures in order to generate proper starting structures for subsequent geometry optimization. We then determined the highest annealing temperature to use for a given cluster based on two considerations: the temperature is (i) not too high to cause the cluster to dissociate into smaller clusters and individual atoms and (ii) high enough to allow different geometrical configurations to be explored. We ran each simulation for 3 ps with a time step of 1 fs, which is comparable to previous studies [38, 42,

45] and is deemed sufficient for exploring the energy landscape of each cluster based on simulated annealing. From the 3000 structural configurations of each cluster, we selected a number of configurations that correspond to different minima on the potential energy landscape of the MD run. In some cases, the cation position of the exported cluster (i.e., "dimer/trimer + cation") was manually adjusted to generate new unexplored configurations. Overall, we used this process to generate 5-11 different configurations for each species (except for single cations), ensuring a wide potential energy surface was explored. All the MD simulations were conducted using the *NVT* ensemble, with the temperature being controlled by a Nose-Hoover thermostat [50, 51]. The DNP basis set and PWC functional were used for the MD simulations to save computational cost [38, 52].

The generated 8-11 configurations for each species were then geometry-optimized using an orbital cutoff of 8.0 Å for all atoms, the DNP basis set, and the BLYP functional without pseudopotential, similar to our previous study [38]. The BLYP exchange-correlation functional was adopted here because this functional has been widely used and proven to be effective for silicate-based systems [38, 39, 43, 53]. The convergence thresholds for energy, force, and displacement were set at $1 \times 10^{-6}$ Hartrees, $2 \times 10^{-4}$ Hartrees/Å, and $5 \times 10^{-4}$ Å, respectively. For some configurations, a thermal smearing of 0.02 Hartrees has been used to assist convergence. For each geometry-optimized structural configuration, we have performed vibrational frequency analysis to obtain its Gibbs free energy at 298.15K and, at the same time, ensure that the configuration is located at a local energy minimum. All the simulations (both the *ab initio* MD and DFT geometry optimization) were performed using the DMol3 v7.0 package, which was part of the Accelrys Materials Studio software.

Once the total energy (i.e., the summation of the ground state energy and Gibbs free energy at 298.15 K) of all configurations for a given species (e.g., $[(OH)_3\text{-Si-O-Al-}(OH)_3]^{-1}$) was determined, we selected the lowest energy value (i.e., the most energetically favorable configuration) and used it to estimate the pair-wise interaction energies (or binding energies) $E_b$ between this negatively charged aluminosilicate species and positively charged cations, as illustrated in Eq. (1).

$$E_b = E_t([AS + R]^{n-m}) - (E_t(R^{n+}) + E_t([AS]^{-m})) \qquad (1)$$

where $[AS]^{-m}$ and $R^{n+}$ refer to the aluminosilicate species with a negative charge of $m$ and the cation with a positive charge of $n$, respectively; $E_t(R^{n+})$ is the calculated total energy of the cation $R^{n+}$; $E_t([AS]^{-m})$ is the calculated total energy of the aluminosilicate species $[AS]^{-m}$; and $E_t([AS + R]^{n-m})$ is the total energy of the reaction product, i.e., the combined cluster of the aluminosilicate species and cation $[AS + R]^{n-m}$. A negative binding energy $E_b$ means that the interaction between the aluminosilicate species and the cation is thermodynamically favorable, with a more negative value indicating a stronger interaction.

As mentioned above, some clusters (e.g., $[AS]^{-m}$ species with the $Fe^{3+}$, $Ni^{2+}$, and $Co^{3+}$ cations) required thermal smearing of 0.02 to assist convergence. Hence, we have evaluated the impact of thermal smearing on the binding energy calculation (based on Eq. (1)) for several clusters (e.g., an aluminosilicate dimer (i.e., $[(OH)_3\text{-}Si\text{-}O\text{-}Al\text{-}(OH)_3]^{-1}$) balanced with a $Li^+$, $Na^+$, or $Ca^{2+}$ cation) that do not have convergence problems. The results are shown in Figure 1, where it is clear that the obtained binding energy values for all three clusters remain almost the same (variations smaller than 1%) at a smearing value less than 0.02. This result suggests that the use of thermal smearing at 0.02 to assist convergence with certain clusters should not significantly change the binding energy calculations and alter the findings and conclusions of this study.

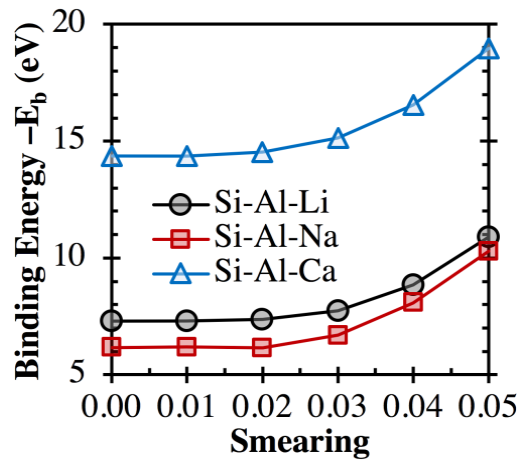

Figure 1. Impact of thermal smearing value on the binding energy between an aluminosilicate dimer (i.e., $[(OH)_3\text{-}Si\text{-}O\text{-}Al\text{-}(OH)_3]^{-1}$) and a $Li^+$, $Na^+$, or $Ca^{2+}$ cation, calculated using Eq. (1).

## 3    Results & Discussion

### 3.1    Optimized structures

Figure 2a-d shows the DFT-optimized cluster structures (i.e., an aluminosilicate dimer $[(OH)_3\text{-}Si\text{-}O\text{-}Al\text{-}(OH)_3]^{-1}$ with a metal cation $M^{n+}$) obtained following the procedures outlined in the Methodology section for a typical monovalent ($Na^+$), bivalent ($Mg^{2+}$), trivalent ($Fe^{3+}$) and tetravalent ($Ti^{4+}$) cation, respectively. The metal cations are seen to form metal-oxygen (M-O) bonds with oxygen atoms in both the silica and alumina tetrahedra. The distance values in Figure 2 show that the M-O bonds formed with oxygen atoms in alumina tetrahedra are shorter than those in silica tetrahedra. Taking $Na^+$ for example (Figure 2a), the average Na-O(Al) distance (~2.40 Å) is about 0.13 Å shorter than the average Na-O(Si) distance (~2.53 Å), which is consistent with the trend seen in previous DFT calculations on Na-exchanged zeolitic frameworks [54]. This observation is also generally true for other cations in Figure S1 of the Supporting Information (optimized structures for other investigated $(OH)_3\text{-}Si\text{-}O\text{-}Al\text{-}(OH)_3 \cdot M^{n+}$ species), which can be attributed to the overall negative charge of the alumina tetrahedra (i.e., $[Al(O_{1/2})_4]^{-1}$), leading to higher attractive forces with the positively charged cations and hence shorter M-O(Al) bond distances (compared with M-O(Si) bonds). These attractive forces pull the oxygen atoms away from the connected Si and Al atoms, leading to longer Si-O(M) (~1.68-1.78 Å) and Al-O(M) (~1.80-1.87 Å) distances compared with other Si-O (~1.61-1.65 Å) and Al-O (~1.69-1.79 Å) bonds away from the metal cations, as clearly seen in Figure 2 and Figure S1 of Supporting Information.

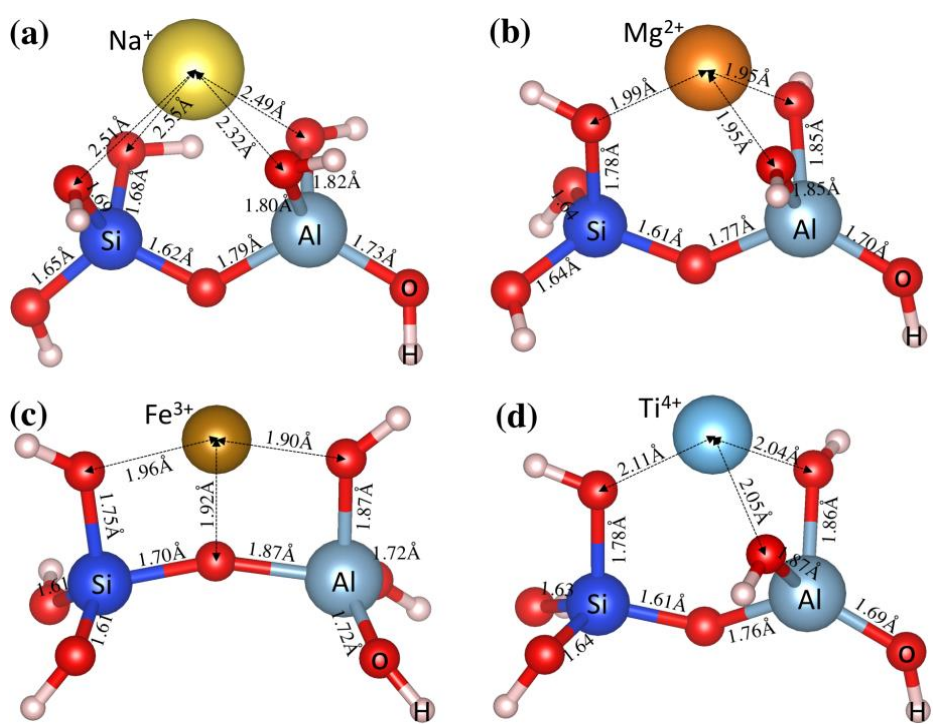

Figure 2. DFT-optimized aluminosilicate dimer (i.e., $[(OH)_3\text{-Si-O-Al-}(OH)_3]^{-1}$) charge balanced by a typical (a) monovalent ($Na^+$), (b) bivalent ($Mg^{2+}$), (c) trivalent ($Fe^{3+}$), and (d) tetravalent ($Ti^{4+}$) cation. Also shown are the interatomic bond distances (in Å).

Figure 3a-d shows the optimized structures for the aluminosilicate trimer (i.e., $[(OH)_3\text{-Al-O-Si-O(-}(OH)_2)\text{-Al-}(OH)_3]^{-2}$) charge-balanced by several typical cations (i.e., $Ca^{2+}$, $Cr^{3+}$, $[CrOH]^{2+}$ and $Ti^{4+}$), with those for the other investigated cations given in Figure S2 of Supporting Information. Figures 3 and S2 show that the metal cations also form M-O bonds with oxygen atoms in both the silica and alumina tetrahedra; however, there appears to be a preferential formation of M-O bonds with bridging oxygen (BO, defined as the oxygen connected to two silica and alumina tetrahedra), rather than NBO in the silica tetrahedra. Furthermore, the M-O bonds formed with BO (e.g., ~2.36 Å for Ca-BO in Figure 3a) are seen to be generally longer than those formed with NBO (e.g., ~2.26 Å for Ca-NBO in Figure 3a), which is consistent with observations in silicate glasses [25]. Another observation from Figures 3 and S2 for the aluminosilicate trimer that is consistent with the aluminosilicate dimer (Figures 2 and S1) is the longer Si-O(M) (~1.65-1.74 Å) and Al-O(M) (~1.81-1.88 Å) bonds compared with the other Si-O (~1.62-1.65 Å) and Al-O (~1.67-1.76 Å) bonds due to the strong interaction between the oxygen atoms and the metal cations (which pull the oxygen atoms away from the Si and Al atoms).

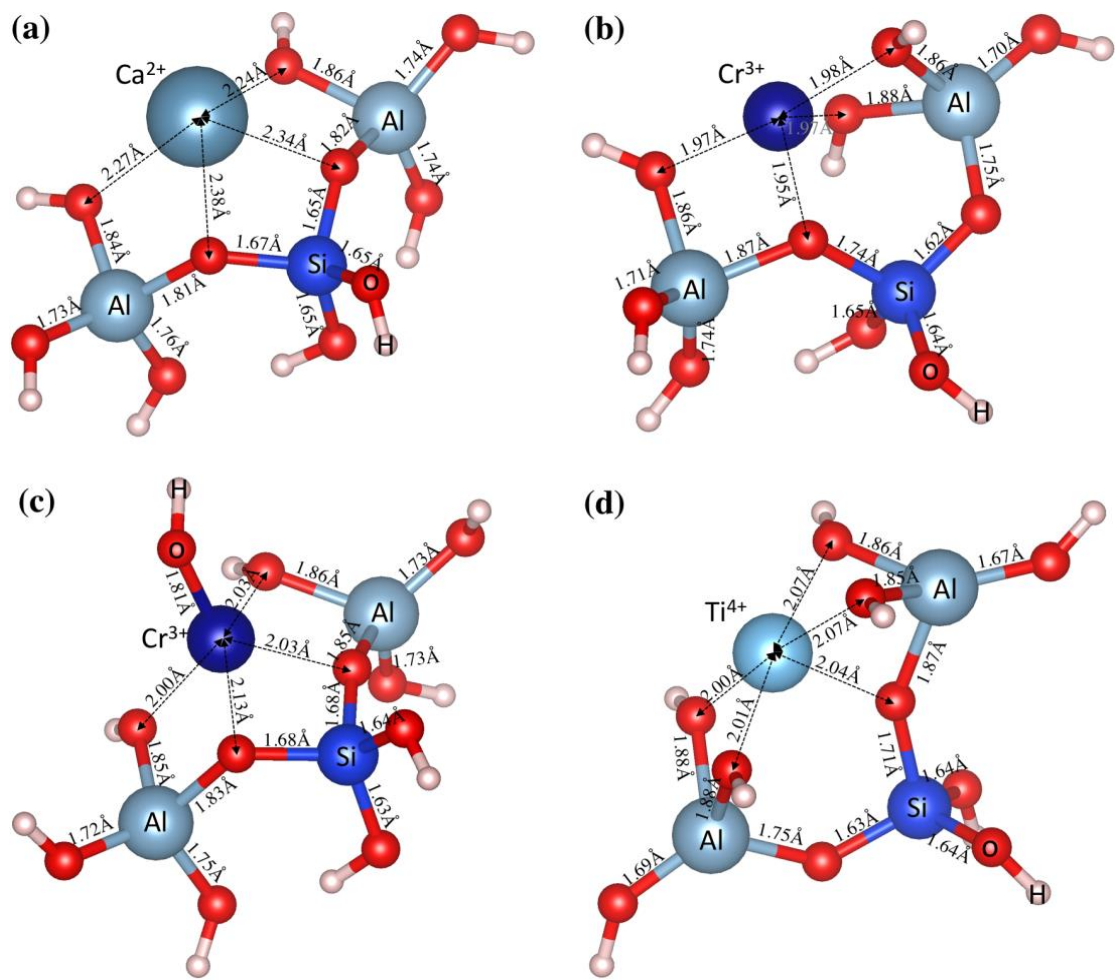

Figure 3. DFT-optimized aluminosilicate trimer (i.e., $[(OH)_3\text{-Al-O-Si-O(-}(OH)_2)\text{-Al-}(OH)_3]^{-2}$), charge-balanced by a typical (a) divalent cation ($Ca^{2+}$), (b) trivalent cation ($Cr^{3+}$), (c) bivalent cluster ($[Cr(OH)]^{2+}$), and (d) tetravalent ($Ti^{4+}$) cation. Also shown are the interatomic bond distances (in Å).

The average interatomic distances for all investigated aluminosilicate clusters are summarized in Table 1 and compared with those reported in the literature for aluminosilicate glasses and/or clusters. The results show that the average Si-O and Al-O distances are around 1.65-1.67 Å and 1.78-1.79 Å, respectively, which are slightly longer than those reported in aluminosilicate glasses (Si-O: ~1.60-1.64 Å [21, 25, 27, 55] and Al-O: ~1.72-1.77 Å [21, 55]). Nevertheless, they are comparable to Si-O and Al-O distances in aluminosilicate clusters reported in previous DFT calculations (Si-O: ~1.64-1.67 Å and Al-O: ~1.76-1.79 Å) [56, 57]. The O-H bond distance varies only slightly at around 0.97-0.99 Å, regardless of the type of chemical complex and cation, where the O-H distance values are comparable to previous DFT calculations on silicate dimers and trimers (~0.95-0.97 Å) [58]. A comparison of the average M-O bond distances in Table 1 shows that the values obtained here in the small aluminosilicate clusters are generally consistent with values reported in the literature on silicate-based glasses and/or clusters. For example, the average Li/Na/K-O distances in our clusters are 1.97, 2.47, and 2.81 Å, respectively, which are within the range reported for silicate glasses in the literature (1.94-2.26, 2.25-2.66, and 2.60-3.12 Å, respectively). The largest deviation is seen for $Ti^{4+}$-O and $Cu^{+}$-O, where our DFT-optimized clusters give average $Ti^{4+}$-O and $Cu^{+}$-O distances of ~2.04-2.07 and ~2.08 Å, slightly larger than those reported in silicate glasses (1.85-1.93 and 1.79-1.91 Å, respectively). Minor deviations (~0.05 Å) between DFT-derived bond distances and literature data can be observed for $Zn^{2+}$-O, and $Fe^{3+}$-O bond distances (Table 1). These differences may be partially caused by the differences in the coordination states and local atomic arrangements of the cations in the clusters studied here and those in silicate glasses.

Table 1. Average interatomic distances for the DFT-optimized cluster structures (see Figures 2-3 and Figures S1-S2 of Supporting Information). Also given in the table are literature data (both experiments and simulations) for metal-oxygen (M-O) bond distances in silicate glasses, minerals and clusters.

| Type of interacting cation or cluster | Interatomic distance (Å) | | | | | | | | | |
|---|---|---|---|---|---|---|---|---|---|---|
| | Aluminosilicate dimer or trimer | | | | | | | Literature experiments | Literature simulations |
| | [(OH)$_3$-Si-O-Al-(OH)$_3$]$^{-1}$ | | | | [(OH)$_3$-Al-O-(OH)$_2$-Si-O-Al-(OH)$_3$]$^{-2}$ | | | | | |
| | $r$(Si-O) | $r$(Al-O) | $r$(O-H) | $r$(M-O) | $r$(Si-O) | $r$(Al-O) | $r$(O-H) | $r$(M-O) | $r$(M-O) | $r$(M-O) |
| H$_3$O$^+$ | 1.657 | 1.794 | 0.977 | **1.023** | n.d. | n.d. | n.d. | **n.d.** | | |
| Li$^+$ | 1.659 | 1.784 | 0.975 | **1.971** | n.d. | n.d. | n.d. | **n.d.** | 1.96-2.24 [59]; 1.94-2.26 [60]; | 1.90 [61]; 1.92-1.97 [25]; |
| Na$^+$ | 1.660 | 1.787 | 0.975 | **2.467** | n.d. | n.d. | n.d. | **n.d.** | 2.30-2.59 [59]; 2.30-2.66 [60]; 2.3-2.62 [55, 62, 63]; | 2.25 [61]; 2.26-3.28 [6]; 2.28 [25]; 2.30-2.39 [22]; |
| K$^+$ | 1.660 | 1.786 | 0.975 | **2.805** | n.d. | n.d. | n.d. | **n.d.** | 2.6-2.7 [55] | 2.60 [61]; 2.72-3.12 [6]; 2.67-2.81 [22]; |
| Cu$^+$ | 1.662 | 1.782 | 0.976 | **2.080** | n.d. | n.d. | n.d. | **n.d.** | 1.79-1.84 [64]; 1.85-1.87 [65]; 1.84-1.91 [66] | |
| Cu$^{2+}$ | 1.661 | 1.782 | 0.975 | **1.912** | 1.656 | 1.785 | 0.974 | **2.159** | | |
| [CuOH]$^+$ | 1.654 | 1.780 | 0.974 | **2.073** | n.d. | n.d. | n.d. | **n.d.** | 1.89-2.23 [64]; 1.95-2.38 [65] | 1.87-2.02 [67] |
| Zn$^{2+}$ | 1.663 | 1.791 | 0.975 | **2.041** | 1.660 | 1.793 | 0.975 | **2.058** | 1.94-1.95 [68]; 1.94-1.97 (MOF) [69]; 1.95-1.99 [70] | 1.96-1.97 (MOF) [69]; 1.93-1.96 [70] |
| Ni$^{2+}$ | 1.660 | 1.785 | 0.974 | **2.014** | 1.659 | 1.786 | 0.974 | **2.074** | 1.98-2.08 [71]; 2.01-2.04 [24] | 1.92-2.09 [72] |
| Ca$^{2+}$ | 1.663 | 1.790 | 0.973 | **2.240** | 1.658 | 1.787 | 0.974 | **2.307** | 2.38 [59], 2.36-2.74 [60]; 2.34-2.36 [27]; 2.25-2.39 [55]; 2.44-2.51 [24] | 2.40-2.61 [6]; 2.35-2.42 [27]; 2.42-2.43 [21]; 2.30-2.39 [55] |
| [CaOH]$^+$ | 1.661 | 1.784 | 0.971 | **2.289** | n.d. | n.d. | n.d. | **n.d.** | | |
| Mg$^{2+}$ | 1.666 | 1.793 | 0.974 | **1.965** | 1.660 | 1.794 | 0.973 | **2.000** | 2.00 [27, 55]; 2.06 [24] | 1.98-2.13 [6]; 2.02-2.03 [27]; 2.03-2.04 [21]; 2.00-2.05 [55] |
| [MgOH]$^+$ | 1.661 | 1.787 | 0.971 | **1.998** | n.d. | n.d. | n.d. | **n.d.** | | |
| Fe$^{2+}$ | 1.660 | 1.789 | 0.974 | **2.032** | 1.662 | 1.790 | 0.974 | **1.986** | 2.01-2.08 [55] | 2.03-2.04 [55] |
| Co$^{2+}$ | 1.660 | 1.789 | 0.974 | **1.969** | 1.661 | 1.792 | 0.975 | **1.976** | 1.95-2.17 [73]; 2.00-2.02 [24] | 2.05-2.12 [73] |
| Ti$^{2+}$ | 1.666 | 1.795 | 0.974 | **2.067** | 1.658 | 1.791 | 0.975 | 2.110 | | |
| Co$^{3+}$ | 1.663 | 1.792 | 0.979 | **1.930** | 1.658 | 1.793 | 0.978 | 1.965 | | |
| [CoOH]$^{2+}$ | n.d. | n.d. | n.d. | **n.d.** | 1.655 | 1.786 | 0.978 | 1.915 | | 1.88 [74] |
| Fe$^{3+}$ | 1.665 | 1.796 | 0.979 | **1.926** | 1.659 | 1.794 | 0.977 | 1.947 | | |
| [FeOH]$^{2+}$ | 1.664 | 1.787 | 0.976 | **1.936** | 1.663 | 1.789 | 0.978 | 1.990 | 1.85 [55]; | 1.86-1.87 [55]; |
| Cr$^{3+}$ | n.d. | n.d. | n.d. | **n.d.** | 1.663 | 1.797 | 0.976 | 1.967 | | |
| [CrOH]$^{2+}$ | n.d. | n.d. | n.d. | **n.d.** | 1.660 | 1.790 | 0.976 | 2.003 | 1.99-2.00 [75]; 1.97 [76] | |
| Ti$^{4+}$ | 1.666 | 1.795 | 0.974 | **2.067** | 1.656 | 1.807 | 0.976 | 2.041 | 1.85-1.89 [55]; 1.85-1.89[22] | 1.80-1.93 [55]; 1.79-1.93[22] |
| Cr$^{6+}$ | n.d. | n.d. | n.d. | **n.d.** | 1.665 | 1.816 | 0.994 | 1.916 | | 1.71-2.00 [77] |

n.d. = not determined.

## 3.2 Binding energies

### 3.2.1 *Monovalent cations*

Based on the total energies of the optimized structures for the individual components (e.g., dimer, trimer, $M^{n+}$, "dimer + $M^{n+}$", and "trimer + $M^{n+}$", as seen in Table S1 of Supporting Information), we have calculated the binding energies between the aluminosilicate dimer/trimer and the different metal cations/clusters using Eq. (1). Figure 4a compares the calculated binding energies between the aluminosilicate dimer (i.e., $[(OH)_3\text{-Si-O-Al-}(OH)_3]^{-1}$) and monovalent cations/clusters, where we see negative binding energies for all cases, suggesting that their interactions are all energetically favorable. This favorability is expected because of the attractive Coulomb interaction between a negatively and a positively charged species. Figure 4b shows that there is a general increase in binding energy (i.e., becomes more negative) as the effective ionic radius of the corresponding monovalent cation decreases. Similar trends have been observed for the interactions of monovalent cations with different chemical species, including (i) guanine and 6-thioguanine tetrads [46], (ii) glutathione [47], (iii) tetraoxa[8]circulene sheet [48], (iv) cubane, cyclohexane and adamantane [49], and (v) methanol [78], where the binding energy (also obtained from DFT calculations in referred studies) increases (i.e., becomes more negative) in the order of $K^+ < Na^+ < Li^+$. Although these chemical species are very different from the aluminosilicate dimer/trimer, their interaction energies with monovalent cations are highly correlated with each other, as illustrated in Figure S3 of Supporting Information. Figure S3 shows that $R^2$ values of 0.94-1.00 are achieved for linear regressions between binding energies obtained here and those reported in the literature for other chemical species [46-49, 78]. The trend observed here (increasing interaction energy in the order of $K^+ < Na^+ < Li^+$) can be attributed to the increase of bond strength (evidenced by bond order) in the order of K-O < Na-O < Li-O, as shown for alkali silicate-based glasses using *ab initio* calculations [26].

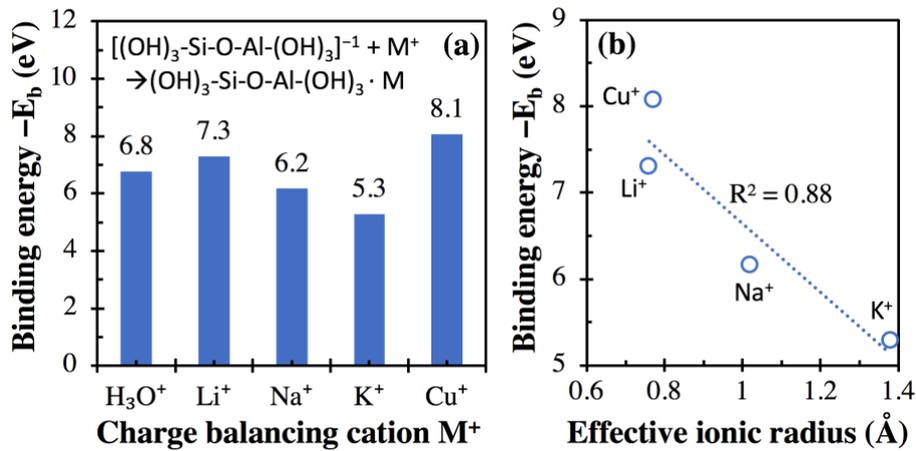

Figure 4. (a) Binding energies (in eV) between the aluminosilicate dimer (i.e., [(OH)$_3$-Si-O-Al-(OH)$_3$]$^{-1}$) and different monovalent cations, calculated using Eq. (1), with the total energies of individual components in Eq. (1) given in Table S1 of Supporting Information. (b) Comparison of binding energies in (a) and the effective ionic radius (for VI-coordinated M$^+$) of the monovalent metal cations. The $R^2$ value for linear regression is also given in (b).

The higher binding energy of Cu$^+$ with the aluminosilicate dimer compared with the other three alkalis M$^+$ (i.e., Na$^+$, K$^+$, and Li$^+$) could be used to explain the observed trend of Cu$^+$⇌M$^+$ exchange in alkali aluminosilicate glasses (i.e., 20M$_2$O-10Al$_2$O$_3$-70SiO$_2$) in an early study [79], where the extent of Cu$^+$⇌R$^+$ exchange is seen to increase in the order of Li$^+$ < Na$^+$ < K$^+$. The occurrence of this Cu$^+$⇌R$^+$ exchange can be partially attributed to the higher binding energy of Cu$^+$ (or bond strength) with the negatively charged aluminosilicate network than the other three alkalis, which promotes the exchange. The trend in the extent of Cu$^+$⇌R$^+$ exchange (Li$^+$ < Na$^+$ < K$^+$) can be attributed to the increasing difference in binding energy between Cu$^+$ and R$^+$ in the order of Li$^+$ < Na$^+$ < K$^+$, which gives an exchange driving force in the order of Li$^+$ < Na$^+$ < K$^+$.

Another observation from Figure 4a is that the binding energy of H$_3$O$^+$ with the aluminosilicate dimer is seen to be higher than K$^+$ and Na$^+$, yet lower than Li$^+$, which can be used to explain the observations in a recent experimental study on alkali aluminosilicate glasses (30M$_2$O-10Al$_2$O$_3$-60SiO$_2$) [80]. Corrosion experiments in this study [80] showed no obvious H$^+$ ⇌ Li$^+$ exchange for the lithium aluminosilicate glass, suggesting that this glass is resistant to moisture attack, likely due to the higher binding energy of Li$^+$ with the aluminosilicate network than H$_3$O$^+$. In contrast, obvious H$^+$ ⇌ Na$^+$ and H$^+$ ⇌ K$^+$ exchange was observed in the sodium- and potassium-containing glasses to a hydrogen penetration depth of 0.4 and 3 μm, respectively, within the same experimental timeframe. This observation is likely due to the lower binding energies of Na$^+$ and K$^+$ (especially the latter, where the binding energy is the lowest among all monovalent cations considered here) with the aluminosilicate network compared to H$_3$O$^+$, as illustrated in Figure 4a.

### 3.2.2 *Divalent cations*

The binding energies of different divalent cations with the aluminosilicate dimer (i.e., [(OH)$_3$-Si-O-Al-(OH)$_3$]$^{-1}$) and trimer (i.e., [(OH)$_3$-Al-O-(OH)$_2$-Si-O-Al-(OH)$_3$]$^{-2}$) are compared in Figure 5a, which shows that the interaction energy with the trimer is consistently higher than those with the dimer by ~7.6-8.2 eV. This is expected given that the trimer has a higher negative charge (−2) than the dimer (−1) and hence exhibits stronger attractive Coulomb interactions with the positively charged divalent cations. A comparison of the energy values in Figures 4a and 5a shows that the divalent cations exhibit higher binding energies with the dimer (~14.5-22.5 eV) than the monovalent cations (~5.3-8.1 eV). This observation is consistent with previous DFT calculations on other chemical complexes [6, 46-49], where divalent cations (e.g., Ca$^{2+}$ and Mg$^{2+}$) are seen to exhibit higher binding energies than monovalent cations (e.g., K$^+$, Na$^+$, and Li$^+$). Again, the energy values from this study are highly correlated with those reported in refs. [6, 46-49], regardless of the type of chemical complex that the metal cations interact with, as seen by the high $R^2$ values (0.94-1.00) achieved with linear regressions in Figure S3 of Supporting Information. These high degrees of correlation among binding energies obtained with different chemical complexes suggest that the differences in cationic binding energies for any given complex are mainly controlled by the inherent properties of the cations (e.g., field strength and ionic potential, as will be shown in the subsequent sections). Similar to Figure 4b, we have examined the correlation between the binding energy values (Figure 5a) and effective ionic radii of the divalent cations in Figure 5b, where we see approximate inverse correlations. Nevertheless, the $R^2$ values achieved with linear regressions (0.66-0.70, Figure 5b) are lower than those in Figure 4b for the monovalent cations (0.88).

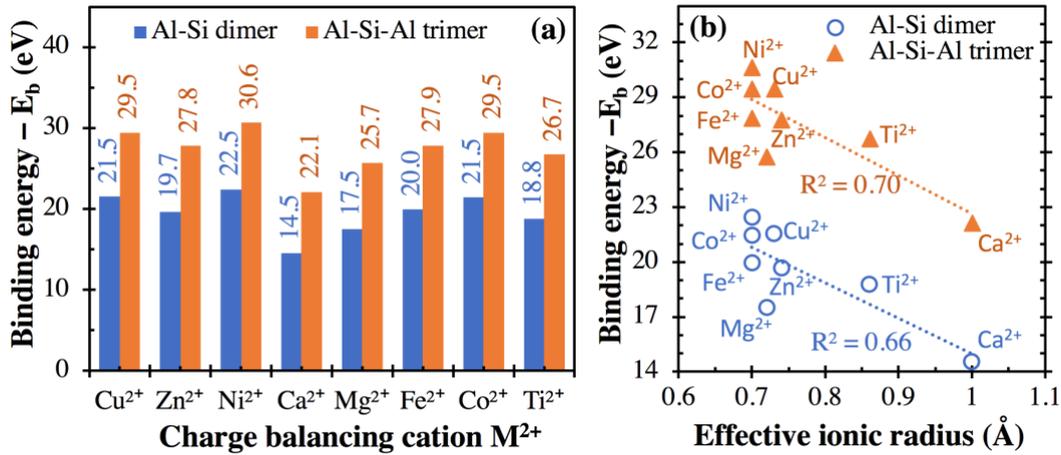

Figure 5. (a) Binding energies between (i) the aluminosilicate dimer (i.e., $[(OH)_3\text{-Si-O-Al-}(OH)_3]^{-1}$) and trimer (i.e., $[(OH)_3\text{-Al-O-}(OH)_2\text{-Si-O-Al-}(OH)_3]^{-2}$) and (ii) different divalent cations, calculated using Eq. (1), with the total energies of individual components in Eq. (1) given in Table S1. (b) Comparison of binding energies in (a) and the effective ionic radius (for VI-coordinated $M^{2+}$) of the divalent metal cations. $R^2$ values for linear regression are also given in (b).

The binding energy results in Figure 5 can also be used to explain observations in many literature investigations. For example, MD simulations on ionic transport in C-A-S-H gel show that the diffusion coefficient of $Ca^{2+}$ is over twenty times smaller than $Na^+$ [81]. The slower diffusion of $Ca^{2+}$, in this case, is likely due to its higher binding strength with the aluminosilicate network (as shown in Figures 4 and 5), which hinders the diffusion of $Ca^{2+}$ in C-A-S-H gel channel (compared with $Na^+$). Furthermore, recent investigations on the chemical durability of alkali/alkaline earth aluminoborate glasses show that the initial glass dissolution rate increases in the order of $Mg^{2+} < Ca^{2+} < Li^+$ [34] and $Mg^{2+} < Li^+ < Na^+$ [33]. These observations can be attributed to the opposite trend in the ability of these cations to stabilize the negatively charged network and to hinder network dissolution/destruction in the order of $Mg^{2+} > Ca^{2+} > Li^+ > Na^+$ (as suggested by the binding energies in Figures 4-5). More recently, the addition of $Mg^{2+}$ [82] and $Cu^{2+}$, and $Co^{2+}$ [83] in AAMs has been shown to reduce AAM leaching (especially leaching of Al species) in sulfuric acid. This observed reduction in dealumination in cation-doped AAMs in refs. [82, 83] may be partially attributed to the higher binding strength of the doped $Mg^{2+}$, $Cu^{2+}$ and $Co^{2+}$ cations that help better stabilize Al compared with $Na^+$ and $Ca^{2+}$ in reference samples (see the binding energy differences in Figures 4-5). This stronger Al stabilization effect of $Mg^{2+}$, $Cu^{2+}$ and $Co^{2+}$ (than $Na^+$ and $Ca^{2+}$) may have two underlying mechanisms: (i) on the one hand, due to their higher binding strength, the doped divalent cations enhance the stability of the negatively charged aluminosilicate network in AAMs (i.e., the main binding gel in the investigated AAMs) upon acid attack, rending it more resistant to destruction; (ii) on the other hand, the doped cations may better stabilize a passivation layer (rich in Si and Al according to SEM-EDX analysis [83]) at the reaction front that slows down diffusion of chemical species.

Figure 6 (and Figure S4 of Supporting Information) compares (i) the binding energy of the aluminosilicate dimer (and trimer) with six divalent cations ($M^{2+}$ = $Ca^{2+}$, $Mg^{2+}$, $Zn^{2+}$, $Fe^{2+}$, $Co^{2+}$, and $Ni^{2+}$) and (ii) the measured log dissolution rate of the corresponding $M_2SiO_4$ and $MCO_3$ minerals at different pHs and the log water exchange rate from the hydration sphere of the corresponding dissolved cation to the surrounding solvent (data extracted from refs. [84, 85]). Although the $R^2$ values achieved using linear regressions are not high (~0.43-0.56), they appear to be inversely correlated, with higher binding energy associated with generally lower dissolution and water exchange rates (Figure 6 and Figure S4). These inverse correlations may be due to the same underlying mechanism: a higher M-O bond strength leads to a higher binding energy in the case of the current study and a higher resistance to M-O bond-breaking for $M_2SiO_4$ and $MCO_3$ mineral dissolution and water exchange. Nevertheless, we note that the mineral dissolution process is highly complex, and measured dissolution rates may be influenced by other factors [35] in addition to M-O bond strength.

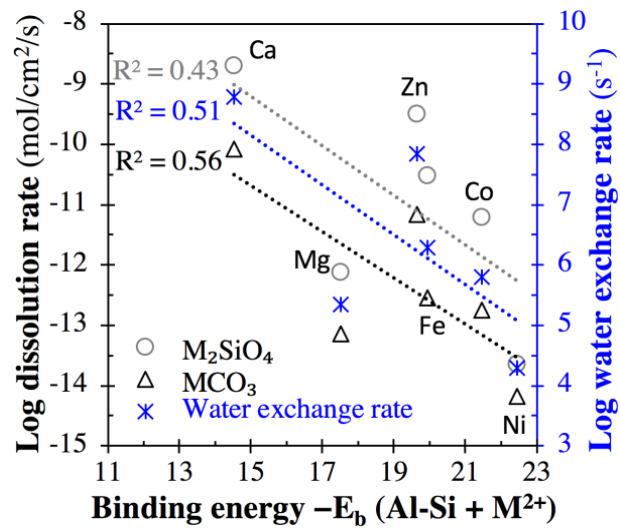

Figure 6. Comparison of the binding energy between the aluminosilicate dimer and divalent cation $M^{2+}$ ($M^{2+}$ = $Ca^{2+}$, $Mg^{2+}$, $Zn^{2+}$, $Fe^{2+}$, $Co^{2+}$, and $Ni^{2+}$) with (i) the log dissolution rate (in mol/cm²/s) of $M_2SiO_4$ minerals at pH = 2 and 25 °C and $MCO_3$ minerals at 5 <pH < 8 and 25 °C (left axis), and (ii) the log rate constant for water exchange from the hydration sphere of the dissolved cation to the surrounding solvent (right axis). Rate data of $M_2SiO_4$ minerals are from ref. [84], while the rate data of $MCO_3$ minerals and water exchange are from ref. [85]. $R^2$ values for linear regression are also given.

### 3.2.3 High valent cations

The calculated binding energy values for the high valent cations (i.e., $Co^{3+}$, $Cr^{3+}$, $Fe^{3+}$, $Ti^{4+}$ and $Cr^{6+}$) are presented in Figure 7, which shows higher interaction energies with the trimer than with the dimer, consistent with the divalent cations in Figure 5a. Furthermore, these binding energy values are considerably higher (i.e., more negative) than those for the monovalent and divalent cations given in Figures 4a and 5a, respectively. For example, the binding energies of $Fe^{3+}$ (39.6 and 53.0 eV with the dimer and trimer, respectively) are considerably higher than those of $Fe^{2+}$ (20.0 and 27.9 eV, Figure 5a). This may explain the higher strength of $Fe^{3+}$-O bonds compared to $Fe^{2+}$-O bonds [22] and the lower dissolution rate of Fe-containing minerals under oxidative conditions compared with reductive conditions, as has been reported in the literature [35].

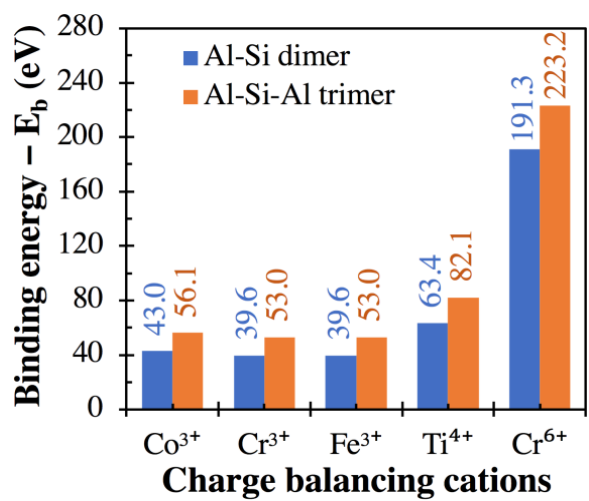

Figure 7. Binding energies between (i) the aluminosilicate dimer (i.e., $[(OH)_3\text{-Si-O-Al-}(OH)_3]^{-1}$) and trimer (i.e., $[(OH)_3\text{-Al-O-}(OH)_2\text{-Si-O-Al-}(OH)_3]^{-2}$) and (ii) different high valent cations, calculated using Eq. (1), with the total energies of individual components in Eq. (1) given in Table S1.

In fact, as seen in Figure 8a, there appears to be a general trend of increasing binding energy with both the aluminosilicate dimer and trimer as the charge of the cation/cluster increases (all the cations and cationic clusters containing $OH^{-1}$ have been included in the figure). These binding energy values are plotted in Figure 8b against the effective ionic radius of the corresponding cations, which shows that they are generally inversely correlated, as already seen in Figures 4b and 5b for monovalent and divalent cations, respectively. Due to these opposite correlations seen in Figures 8a and 8b, we have plotted in Figure 9a the binding energy values as a function of cation charge/ionic radii, which is defined as the ionic potential (IP) of the cation introduced by Cartledge [86] to describe to what extent the cations are electrostatically attracted by oppositely charged ions. Here, we used the effective ionic radii tabulated by Shannon [87] to calculate the ionic potential of each studied cation (see the values in Table 2). It is clear from Figure 9a that the cationic binding energies with the aluminosilicate dimer/trimer are positively correlated with the ionic potential of the cations, and their correlations can be accurately captured using 2$^{nd}$ order polynomial functions, as evidenced by the high $R^2$ values achieved for regressions (0.99-1.00). Previous studies have also attempted to connect the ionic potential of cations to their binding energies with organic species (e.g., glutathione [47] and calix[2]furano[2]pyrrole [88]), as well as to other material properties (e.g., formation enthalpies [89, 90] and cation discharge potential [86]).

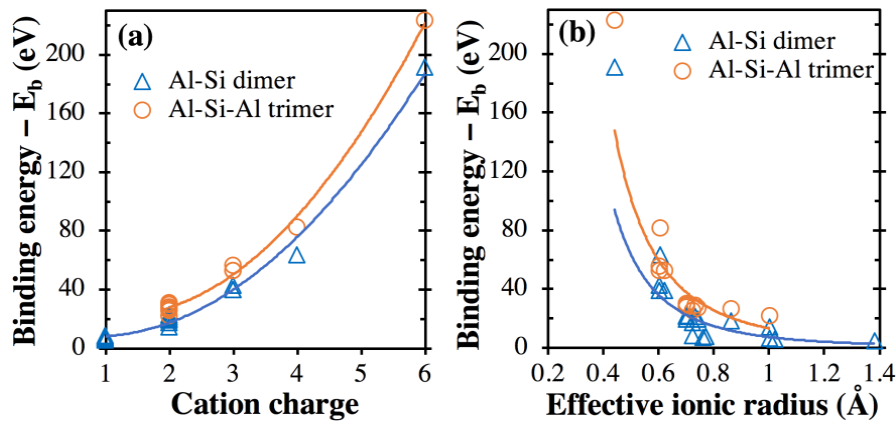

Figure 8. Comparison of binding energies of all the metal cations (including all the metal cation+OH clusters, e.g., [Ca(OH)]$^+$) with the aluminosilicate dimer/trimer and (a) the cation/cluster charge (b) the cation effective ionic radius. In the case of cation+OH clusters (e.g., [Ca(OH)]$^+$), the effective ionic radius of the corresponding cation (i.e., Ca$^{2+}$) is used. The lines are given to guide the eye.

In the glass community, one important term introduced by Dietzel [91] to characterize the effect of a single cation on oxide glasses is the cationic field strength ($F$) parameter, which is given by Eq. (2) [92]:

$$F = \frac{z_C}{(r_C + r_{O^{2-}})^2} \qquad (2)$$

where $Z_C$ is the charge of the cation; $r_C$ and $r_{O^{2-}}$ are the ionic radii of the cation and oxygen anion, respectively. This field strength parameter has been widely used to classify ions as network modifiers ($F \approx 0.1$-$0.4$), network formers ($F \approx 1.3$-$2.1$), and intermediates ($F \approx 0.5$-$1$) [92]. Again, here we have used the effective ionic radii from Shannon [87] to calculate the field strength of each cation (see the values in Table 2), where the results are presented in Figure 9b as a function of the calculated binding energies with the aluminosilicate dimer and trimer. The binding energy values are seen to be positively correlated with the cationic field strength, and the correlations can be well captured by 2$^{nd}$ order polynomial functions ($R^2$ values of 0.99-1.00), similar to the case of the ionic potential (Figure 9a). In fact, as shown in Figure S5, the ionic potential and field strength of the cations are positively and linearly correlated with an $R^2$ value of 0.99 for a linear regression. In the glass literature, many studies have attempted to use cationic field strength to draw connections with the properties of silicate-based glasses, including effective ionic diffusion [32], glass transition temperature [28, 30, 31], and network connectivity (e.g., Al and B coordination) [28, 30, 31].

Table 2. Summary of effective ionic radii ($r_C$) of the studied cations (obtained from Shannon [87]), along with calculated ionic potential ($Z_C/r_C$) and field strength ($Z_C/(r_C + r_{O^{2-}})^2$).

| Cation | Cation effective radius $r_C$ (Å) | Oxygen effective radius $r_{O^{2-}}$ (Å) | Cationic charge $Z_C$ | Ionic potential (IP) (Å$^{-1}$) | Field strength (F) (Å$^{-2}$) |
|---|---|---|---|---|---|
| Li$^+$ | 0.76 | 1.40 | 1 | 1.32 | 0.21 |
| Na$^+$ | 1.02 | 1.40 | 1 | 0.98 | 0.17 |
| K$^+$ | 1.38 | 1.40 | 1 | 0.72 | 0.13 |
| Cu$^+$ | 0.77 | 1.40 | 1 | 1.30 | 0.21 |
| Cu$^{2+}$ | 0.73 | 1.40 | 2 | 2.74 | 0.44 |
| Zn$^{2+}$ | 0.74 | 1.40 | 2 | 2.70 | 0.44 |
| Ni$^{2+}$ | 0.70 | 1.40 | 2 | 2.86 | 0.45 |
| Ca$^{2+}$ | 1.00 | 1.40 | 2 | 2.00 | 0.35 |
| Mg$^{2+}$ | 0.72 | 1.40 | 2 | 2.78 | 0.44 |
| Fe$^{2+}$ | 0.70 | 1.40 | 2 | 2.86 | 0.45 |
| Co$^{2+}$ | 0.70 | 1.40 | 2 | 2.86 | 0.45 |
| Ti$^{2+}$ | 0.86 | 1.40 | 2 | 2.33 | 0.39 |
| Co$^{3+}$ | 0.60 | 1.40 | 3 | 5.00 | 0.75 |
| Cr$^{3+}$ | 0.62 | 1.40 | 3 | 4.84 | 0.74 |
| $Fe^{3+}$ | 0.60 | 1.40 | 3 | 5.00 | 0.75 |
| $Ti^{4+}$ | 0.61 | 1.40 | 4 | 6.61 | 1.00 |
| $Cr^{6+}$ | 0.44 | 1.40 | 6 | 13.64 | 1.77 |
| $[CuOH]^+$ | 0.77 | 1.40 | 1 | 1.30 | 0.21 |
| $[CaOH]^+$ | 1.00 | 1.40 | 1 | 1.00 | 0.17 |
| $[MgOH]^+$ | 0.72 | 1.40 | 1 | 1.39 | 0.22 |
| $[CoOH]^{2+}$ | 0.70 | 1.40 | 2 | 2.86 | 0.45 |
| $[CrOH]^{2+}$ | 0.62 | 1.40 | 2 | 3.23 | 0.49 |
| $[FeOH]^{2+}$ | 0.60 | 1.40 | 2 | 3.33 | 0.50 |

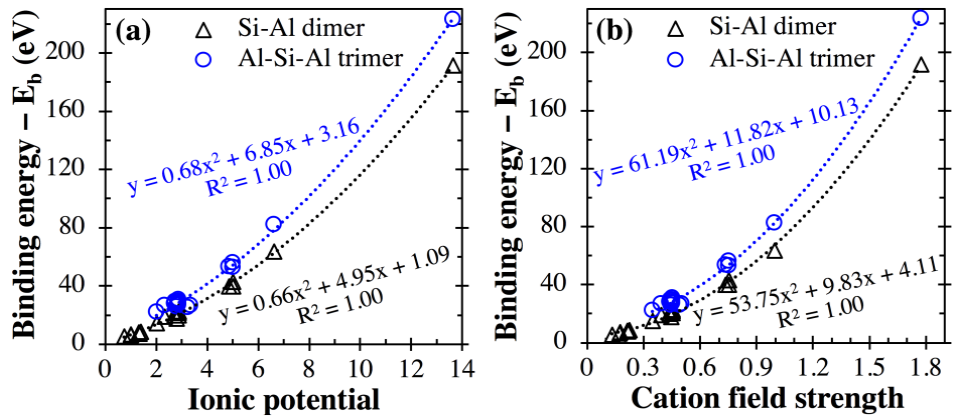

Figure 9. Comparison of the cation binding energies (including cation+OH clusters) with the aluminosilicate dimer and trimer with (a) the ionic potential (i.e., charge/effective ionic radius) and (b) the field strength (see Eq. (2)) of the metal cations. Data have been fitted using 2nd order polynomial functions, where the equations of best fit and corresponding $R^2$ values are provided.

With these simple polynomial functions in Figure 9 (also given in Eqs. (3)-(4)) obtained using regression, we can now provide approximate estimates of the binding energies of different cations with the aluminosilicate dimer/trimer by using either the ionic potential or the field strength of the cations, where the ionic potential and field strength of the cations can be readily estimated from well-tabulated ionic radii available in the literature for a wide range of cations beyond those studied here [87]. For example, we can estimate binding energies of different cations with the aluminosilicate dimer, including cations beyond those included in the DFT calculations (e.g., $Rb^+$, $Cs^+$, $Be^{2+}$, $Ba^{2+}$, $Sr^{2+}$, $Cd^{2+}$, $Pb^{2+}$, and $Al^{3+}$), by using Eqs. (3)-(4).

$$-E_b = 0.66 \times IP^2 + 4.95 \times IP + 1.09 \qquad (3)$$

$$-E_b = 53.75 \times F^2 + 9.83 \times F + 4.11 \tag{4}$$

The calculated cationic binding energies with the aluminosilicate dimer (values given in Table S2 of Supporting Information) are plotted in Figure 10 against the corresponding cationic binding energies with different organic species reported in the literature (calix[2]furano[2]pyrrole ($C_{20}N_2O_2H_9$) [88], β-cyclodextrin ($C_{42}H_{70}O_{35}$) [93] and cucurbit[6]uril ($C_{36}H_{36}N_{24}O_{12}$) [94]), where the binding energies from refs. [88, 93, 94] have been obtained using DFT calculations. It is clear from Figure 10 that the cationic binding energies estimated using Eqs. (3)-(4) for the aluminosilicate dimer are positively and linearly correlated with the DFT-derived binding energies for all three types of organic species. In spite of the large difference in the types of interacting chemical complexes (see the atomic structures of the three organic species in refs. [88, 93, 94]), these linear correlations ($R^2$ values of 0.95-0.99 for linear regressions) in Figure 10 suggest that the relative binding energy values estimated using Eqs. (3)-(4) are reasonable, with the overall trend for a range of cations being correctly captured. Nevertheless, we also see that the estimated values from Eqs. (3)-(4) are generally more than 30% higher than those from DFT calculations. This discrepancy may be partially attributed to the difference in the charge of the organic complex (neutral) compared with the Al-Si dimer (–1). This is supported by Figure 9, where the interaction energies with the more negatively charged trimer (–2) are consistently higher (~30-50% higher) than the same cation with the dimer. The linear correlations in Figure 10 and Figure S3 also demonstrate the governing effect that the ionic potential and field strength of cations have on the binding energies with a given chemical species.

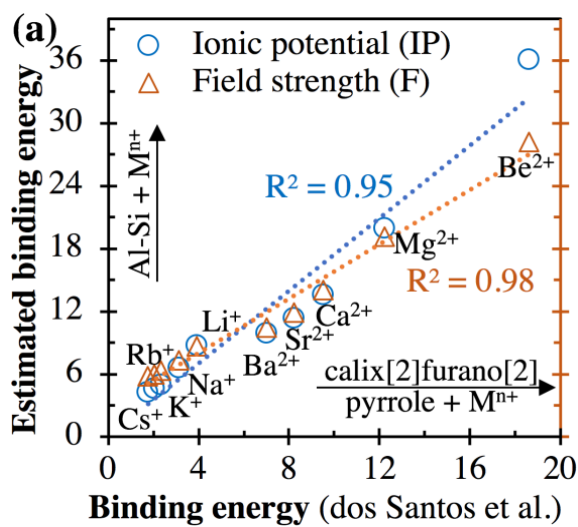

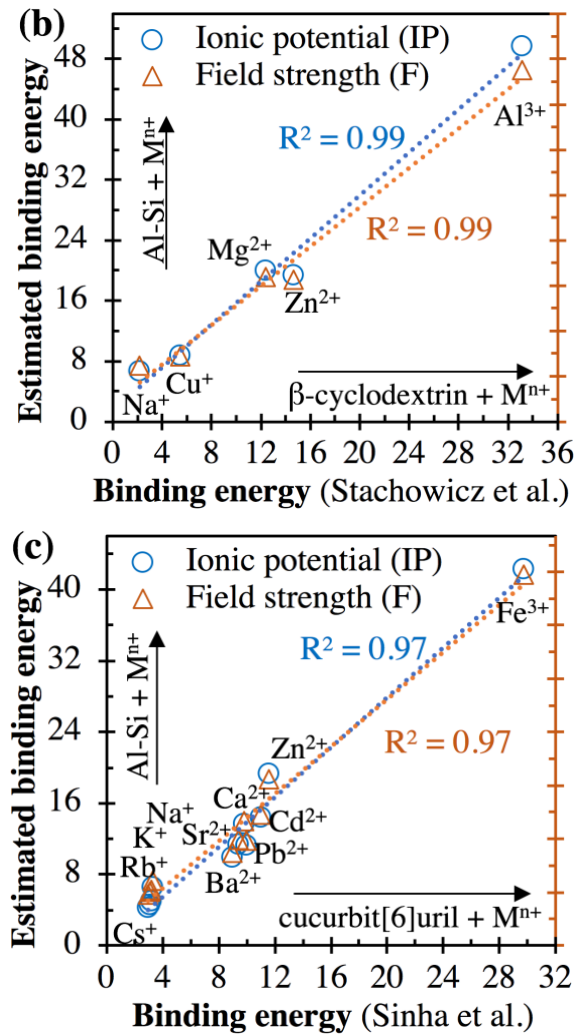

Figure 10. Comparison of the cationic binding energies with the aluminosilicate dimer ($[(OH)_3$-$Si$-$O$-$Al$-$(OH)_3]^{-1}$) estimated using Eqs. (3)-(4) and the DFT-derived binding energies of same cations interacting with three organic species: (a) calix[2]furano[2]pyrrole ($C_{20}N_2O_2H_9$) [88], (b) β-cyclodextrin ($C_{42}H_{70}O_{35}$) [93] and (c) cucurbit[6]uril ($C_{36}H_{36}N_{24}O_{12}$) [94]. All energies are in units of eV. The range of cation types studied in each of the three literature studies is provided, along with the $R^2$ values achieved for linear regressions.

### 3.3 Broader impact and limitations

The interactions of positively charged metal cations with negatively charged aluminosilicate networks are important to many aluminosilicate-based materials (e.g., sustainable cements and aluminosilicate glasses) and their associated applications (e.g., building and construction, waste encapsulation, and durable glasses). Here, we probe this interaction using simple model systems, i.e., calculating the pair-wise interaction energies between aluminosilicate dimer/trimer and different metal cations/clusters using DFT calculations. By covering a wide range of cations, we reveal that simple 2$^{nd}$ order polynomial functions can be used to estimate the binding energy values based on ionic potential or field strength, which can then be estimated using well-tabulated ionic radii available in the literature. With these equations, one can rapidly estimate the binding energies with the aluminosilicate dimer/trimer for a wider range of cations in the periodic table. This presents enormous opportunities for the design and optimization of aluminosilicate-based materials for specific applications, given that there is a strong correlation between the binding energies of cations and their impact on different materials properties (e.g., aluminosilicate glass corrosion, leaching and acid attack of AAMs, ionic transport in AAMs, and mineral dissolution), as seen in the discussion of Section 3.2.

For example, microbial-induced sulfuric acid attack represents a major durability issue for concrete sewer pipelines, requiring an estimated 390 billion dollars in the USA alone over the next 20 years for maintenance and replacement [95]. Recent studies have demonstrated the potential of doping $Mg^{2+}$, $Cu^{2+}$, and $Co^{2+}$ ions to improve the resistance of AAMs to sulfuric acid attack [82, 83]. The results of this study suggest that there are cations (e.g., $Fe^{3+}$ and $Ti^{4+}$) that may better stabilize the aluminosilicate network in AAMs and hence further improve their resistance to acid attack. A wider range of cations can be quickly evaluated for this application using the semi-empirical equations (as given in Figure 9 for the aluminosilicate dimer and trimer) derived here prior to carrying out validation experiments. However, it is noted that the actual performance of elemental doping also depends on the cost of the doped elements and to what extent the doped cations can be incorporated into the AAM structure while not significantly compromising their other properties (e.g., development of strength). Furthermore, similar methods and analysis could be extended to other systems, where binding energy values based on DFT calculations are available, as shown for some organic species [46-49, 78, 88, 93, 94].

Nevertheless, several limitations regarding this work warrant some discussion. First, the interactions between metal cations and the aluminosilicate network are more complex than those with the aluminosilicate dimer/trimer, especially considering the second role of metal cations in aluminosilicates (in addition to charge balancing), i.e., acting as a modifier cation to depolymerize the aluminosilicate network. This impact of depolymerization is opposite to the stabilization effect of cationic charge balancing and hence needs to be considered when using the method and analysis presented here. Nevertheless, this study, along with many others in the literature, has demonstrated the benefits of using DFT calculations and simple model systems to gain physical insight into complex material systems.

Second, although Figure 9 shows that the overall trend of binding energy can be well captured by the ionic potential and field strength of the cations with high $R^2$ values (0.99-1.00), the correlations for the divalent cations alone are much lower ($R^2$ values of 0.65-0.69) as seen in Figure S6 of Supporting Information. Furthermore, similar to Figure S6, we have correlated the ionic potential and field strength of divalent cations with their binding/adsorption energies with several other chemical species reported in the literature (see Figure S7 of Supporting Information). Figure S7 shows that although the binding (adsorption) energies for the divalent cations with 1,10-phenanthroline complexes [96] (silica-disiloxane cluster [97]) are positively correlated with both the ionic potential and field strength of the cations, their levels of correlation are obviously lower than those in Figure 9, but comparable with those shown in Figure S6. One possible contribution to the lower levels of correlation seen for the divalent cations is that the effective ionic radii used to calculate ionic potential and field strength are based on the assumption of VI-coordinated cations. However, this assumption is different from the DFT calculations on the model clusters (as seen in Figures 2 and 3), where the cations are not VI-coordinated; and it also deviates from aluminosilicate glass systems where, for example, $Fe^{2+}$ is mainly V-coordinated, and $Zn^{2+}$ is mainly IV- and V-coordinated [98].

Finally, as shown in Figures 9 and 10, the interaction energies also depend on the type of chemical species (dimer vs. trimer and inorganic vs. organic complex) interacting with the cations. This means that the equations that capture the relationship between the binding energies and the cationic attributes (e.g., IP and field strength) vary among different interacting species (as seen in Figure 9).

## 4  Conclusions

In this study, we employed density functional theory (DFT) calculations to calculate the pair-wise interaction energies (i.e., binding energies) between aluminosilicate dimer/trimer and different metal cations $M^{n+}$ (including $Li^+$, $Na^+$, $K^+$, $Cu^+$, $Cu^{2+}$, $Co^{2+}$, $Zn^{2+}$, $Ni^{2+}$, $Mg^{2+}$, $Ca^{2+}$, $Ti^{2+}$, $Fe^{2+}$, $Fe^{3+}$, $Co^{3+}$, $Cr^{3+}$, $Ti^{4+}$ and $Cr^{6+}$). Comparison with literature data on aluminosilicate glasses shows that the main attributes (e.g., interatomic distances) of DFT-optimized cluster (dimer/trimer + $M^{n+}$) structures are reasonable. The DFT-derived binding energies are seen to increase (i.e., become more negative) as the charge of the metal cation and aluminosilicate increase, whereas these energies decrease (i.e., become less negative) as the radii of the metal cation increase. Comparison with literature data shows that the cationic binding energy can be used to explain many literature observations on the impact of metal cations on the properties of aluminosilicate materials (including aluminosilicate glass corrosion, leaching and acid attack of alkali-activated materials (AAMs), ionic transport in AAMs, and mineral dissolution). These binding energies are shown to be highly correlated ($R^2$ values of 0.94-1.00 for linear regression) with the reported binding energy values in the literature (also obtained using DFT calculations) on the same cations but with different chemical species for interaction (mostly organic complexes), suggesting the presence of certain inherent attributes of the metal cations that control their strength of interaction with a given chemical species.

Analysis of all the DFT-derived binding energies from this study reveals that these energy values can be approximated as a function of two fundamental properties of the metal cations ($R^2$ values of 0.99-1.00 are achieved using regression of 2$^{nd}$ order polynomial function), namely the ionic potential (charge/radii) and field strength (Eq. (2)). This means that the binding energies of a given metal cation with the aluminosilicate dimer/trimer can be readily estimated using simple polynomial functions since both the ionic potential and field strength of the cation can be computed from ionic radii that are well-tabulated in the literature. This is demonstrated for eight cations ($Cs^+$, $Rb^+$, $Sr^{2+}$, $Ba^{2+}$, $Cd^{2+}$, $Pb^{2+}$, $Be^{2+}$, and $Al^{3+}$), where the estimated binding energies using these polynomial functions (Eq. (3)-(4)) are seen to be linearly correlated ($R^2$ values of 0.95-0.99) with DFT-derived interaction energies for different organic species reported in the literature. The differences in the interaction energies among the aluminosilicate dimer and trimer and different organic species with the same cations show that the attribute of the interacting species also has a role to play. The findings in this study present a bottom-up approach (e.g., tailoring the cationic binding energy) toward the design and optimization of sustainable cements and aluminosilicate glasses for specific applications (e.g., improving the resistance of AAM to acid attack, a major durability issue of concrete materials and structures). Similar approaches can be extended to study the binding energies of cations with other chemical species and to derive semi-empirical governing equations that allow rapid estimation of binding energies across a wider range of atoms in the periodic table.

## 5  Supporting Information

1. Optimized dimeric clusters

2. Optimized trimeric clusters
3. Summary of all binding energy values
4. Comparison of binding energies
5. Comparison of binding energy and mineral dissolution rates
6. Comparison of ionic potential and field strength of cations
7. Binding energy values in Figure 10 of the main article
8. Comparison of binding energy and the ionic potential and field strength for divalent cations

# 6 Acknowledgments

This material is based on work supported by ARPA-E under Grant No. 1953-1567.